\documentclass{aa}
\usepackage{natbib}
\bibpunct{(}{)}{;}{a}{}{,}
\usepackage{graphicx}
\usepackage{txfonts}
\begin{document}
   \title{The SW~Sex-type star 2MASS J01074282+4845188: an unusual\\
bright accretion disk with non-steady emission and \\
a hot white dwarf
    \thanks{Based on data collected with telescopes at Rozhen National Astronomical Observatory.}}

%   \subtitle{ }

   \author{T. Khruzina\inst{1} \and D. Dimitrov\inst{2} \and D. Kjurkchieva\inst{3}}

   \offprints{D. Dimitrov}

   \titlerunning{2MASS~J01074282+4845188: unusual bright accretion disk}
   \authorrunning{T. Khruzina et al.}

   \institute{Moscow MV Lomonosov State University, Sternberg Astronomical Institute, Moscow 119991, Russia\\
              \email{kts@sai.msu.ru}
          \and
          Institute of Astronomy and National Astronomical Observatory, Bulgarian Academy of Science, Sofia 1784, Bulgaria\\
          \email{dinko@astro.bas.bg}
          \and
          Shumen university, Shumen 9700, Bulgaria\\
          \email{d.kyurkchieva@shu-bg.net}
          }
   \date{Received 19 September 2012/ Accepted 22 January 2013}

% \abstract{}{}{}{}{}
% 5 {} token are mandatory

  \abstract
   {Cataclysmic variables (CVs) present a short evolutional stage of binary systems. The nova-like stars are rare objects,
especially those with eclipses (only several tens). But precisely
these allow to determine the global parameters of their
configurations and to learn more about the late stage of stellar
evolution.}
   {The light curve solution allows one to determine the global parameters of the newly discovered nova-like eclipsing star 2MASS~
J01074282+4845188 and to estimate the contribution of the different light sources.}
   {We present new photometric and spectral observations of 2MASS~J01074282+4845188. To obtain a light curve solution we used
model of a nova-like star whose emission sources are a white dwarf
surrounded by an accretion disk, a secondary star filling
its Roche lobe, a hot spot and a hot line. The obtained
global parameters are compared with those of the
eclipsing nova-like UX UMa.}
   {2MASS~J01074282+4845188 shows the deepest permanent eclipse among the known nova-like stars. It is reproduced by
covering the very bright accretion disk by the secondary
component. The luminosity of the disk is much bigger than that of
the rest light sources. The determined high temperature of the
disk is typical for that observed during the outbursts of CVs. The
primary of 2MASS~J01074282+4845188 is one of the hottest white
dwarfs in CVs. The temperature of 5090 K of its secondary is also
quite high and more appropriate for a long-period SW Sex star. It
might be explained by the intense heating from the hot white dwarf
and the hot accretion disk of the target.}
   {The high mass accretion rate $\dot{M} = 8\times 10^{-9} M_{\sun}$~yr$^{-1}$, the broad and single-peaked H$\alpha$ emission
profile, and the presence of an S-wave are sure signs for the SW
Sex classification of 2MASS~J01074282+4845188. The obtained flat
temperature distribution along the disk radius as well as the
deviation of the energy distribution from the black-body law are
evidence of the non-steady emission of the disk. It can
be attributed to the low viscosity of the disk matter due to its
unusual high temperature. The close values of the disk temperature
and the parameter $\alpha_{g}$ of 2MASS~J01074282+4845188 and
those of the cataclysmic stars at eruptions might be considered as
an additional argument for the permanent active state of nova-like
stars.}


   \maketitle

\section{Introduction}

Cataclysmic variables (CVs) consist of a white dwarf surrounded by
an accretion disk and a late main sequence star filling
its Roche lobe. The nova-like stars are nonmagnetic cataclysmic
variables that do not truly erupt. It is assumed that they are in
state of permanent eruption due to their high accretion rate.
According to the standard models, the dominant emission sources of
nova-like variables are the accretion disks. The second light
contribution comes from the hot spot at the region where the
accretion stream from the secondary star falls on the
disk. These configurations explain the hump, flickering, and
standstill of their light curves.

Recent gas-dynamical investigations
\citep[][etc.]{bis97,bis98,bis05} revealed that there is a place
of increased energy output of nova-like stars outside the disk.
This region, called \textit{hot line}, is a result of the
interaction of the disk halo and the inter-component envelope with
the gas stream and allows one to fit the whole light curves of the
nova-like stars, including the variability of their out-of-eclipse
parts \citep{khruz01,khruz03}.

The cataclysmic variables present a short evolutional stage of
binary systems. That is why they are rare objects, especially
those with eclipses (only several tens). But precisely these allow
one to determine the global parameters of their
configurations and to learn more about the late stage of stellar
evolution. Therefore, one should not miss the chance to
investigate each cataclysmic variable with an eclipse.

2MASS~J01074282+4845188 = T-And0-10518 =
1SWASP~J010744.41+484458.1 (hereafter J0107) is a star from the
list of 773 eclipsing binary (EB) systems of the TrES database
\citep{devor08}. It has been designated as an ``ambiguous EB''
with coordinates $\alpha = 01^{\rm{h}} 07^{\rm{m}} 44\fs417$,
$\delta = +48\degr 44\arcmin 58\farcs11$ and an orbital period of
$0\fd1935761$.

The recent photometric observations of \citet[][hereafter
Paper~I]{dim12} \defcitealias{dim12}{Paper~I} led to the
conclusion that \citet{devor08} misidentified the variable
T-And0-10518 for its neighbor, a considerably brighter star
(standard star St8 in fig. 2 and table 1 of \citetalias{dim12}) at
distance of $26\arcsec$ from the true variable. The
misidentification is due to the low spatial resolution of the TrES
observations. As a result of the impossibility to separate the
contributions of the close stars, the determined initial
variability amplitude has been almost 11 times smaller than the
true amplitude of the fainter star.

The follow-up observations with a bigger telescope
\citepalias{dim12} than the one used for the TrES survey allowed
us to separate the two close stars, to determine which of them is
the variable, and to obtain the correct variability amplitude of
the precise light curve. It was found that the $VRI$ light curves
of the target have a very deep, asymmetric minimum. The V-shaped
light minimum, the characteristic phase dependence of the color
indices $V-R$ and $V-I$, the pre-eclipse hump in the light curves,
and the flickering led to the conclusion that this star belongs to
the nova-like CVs. The observed broad H$\alpha$
emission profile confirmed the photometric classification. The
detailed analysis of the photometric and spectral data of J0107
led to the suggestion that it belongs to the nova-like CVs 
of SW Sex subtype \citepalias{dim12}.

The goal of our present study is to obtain a common light
curve solution for the data from \citetalias{dim12} and the new
photometric data and to analyze the global parameters of
the newly discovered cataclysmic star J0107.

\section{Rozhen observations and initial analysis}

\subsection{Photometry}

The photometric observations of J0107 at the Rozhen National
Astronomical Observatory (Table \ref{tab:1}) were carried out with
(i) the 60-cm Cassegrain telescope using the FLI PL09000 CCD
camera (3056 $\times$ 3056 pixels, 12 $\mu$m/pixel, field of 27.0
$\times$ 27.0 arcmin with focal reducer); (ii) the 50/70-cm
Schmidt telescope equipped by the CCD camera FLI PL16803 (4096
$\times$ 4096 pixels, 9 $\mu$m/pixel, field of 73.3 $\times$ 73.3
arcmin); (iii) the 2-m RCC telescope with the CCD camera VersArray
512 (512 $\times$ 512 pixels, 24 $\mu$m/pixel, field of 7.6
$\times$ 7.6 arcmin). The average accuracy of the photometric
observations in the filters $B$, $V$, $R$ and $I$ was $0\fm010$,
$0\fm020$, $0\fm010$, and $0\fm008$.

\begin{table*}
\caption{Journal of the Rozhen photometric and spectral
observations} \label{tab:1} \centering
\begin{tabular}{l c c c c c c}
\hline\hline
Date       &Filter & Exp. [s]& N      & phase range  &Telescope& Source \\
\hline
2011 Jan 29& $V,I$ & 120, 120& 67, 67 &  whole cycle & 60-cm  & \citetalias{dim12} \\
2011 Jan 30& $V,I$ & 120, 120& 65, 65 &  whole cycle & 60-cm  & \citetalias{dim12} \\
2011 Jan 31& $R$   & 60      & 222    &  0.86 cycle  & 60-cm  & \citetalias{dim12}\\
2011 Feb 07&$spectra$&300    & 4      &  0.1 cycle   & 2-m    & \citetalias{dim12}\\
2011 March 12&$spectra$&300  & 4      &  half cycle  & 2-m    & new \\
2011 Aug 18& $R$   & 60      & 58     &  half cycle  & 60-cm  & new \\
2011 Nov 04&$spectra$&300    & 2      &  0.05 cycle  & 2-m    & new \\
2011 Nov 11& $B$   & 90      & 261    &more one cycle& 2 m    & new \\
2011 Dec 16& $V$   & 60      & 262    &  whole cycle & Schmidt& new \\
\hline
\end{tabular}
\end{table*}

The new photometric data were reduced in the same way as in
\citetalias{dim12} and the standard stars were the same.

\begin{figure}
\resizebox{0.95\hsize}{!}{\includegraphics{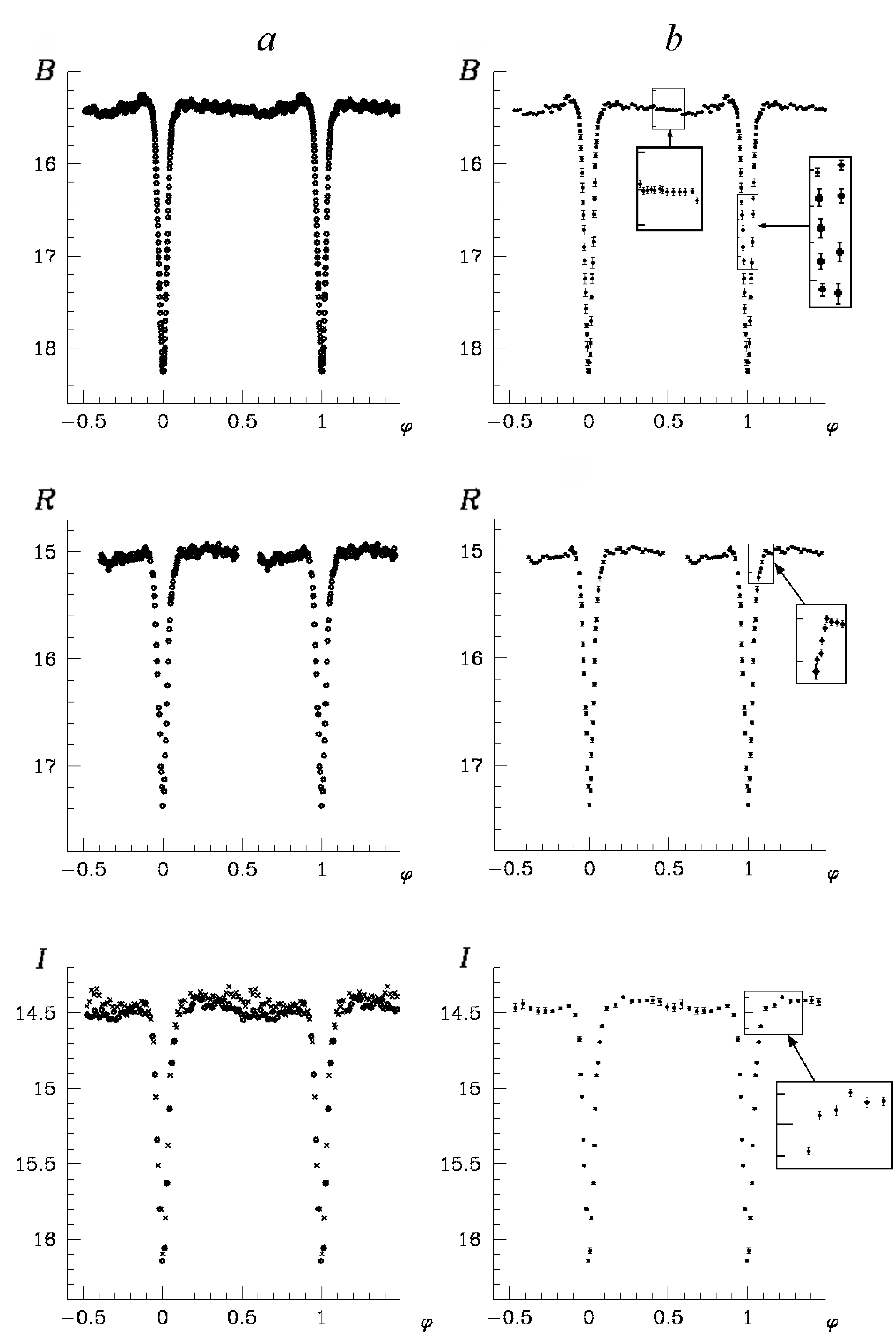}}
\caption{Individual (\textit{a}) and average (\textit{b}) $BRI$
light curves of J0107. The $I$ photometric data are shown by
different symbols: points for 2011 Jan 29 and crosses for 2011 Jan
30. The marked parts of the light curves are also shown on a large
scale for better visibility of the error bars.} \label{fig:1}
\end{figure}

\begin{figure*}
%\sidecaption
\includegraphics[width=0.95\hsize]{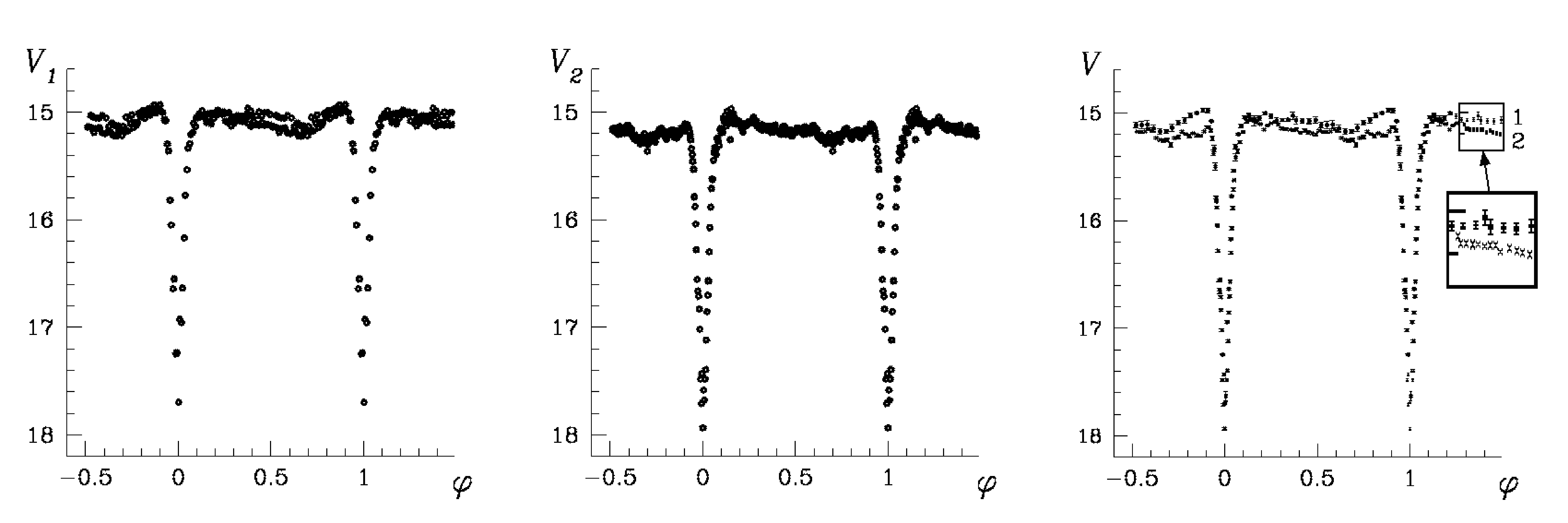}
\caption{From left to right: light curve $V_{1}$ of J0107 from
January 2011; light curve $V_2$ from December 2011; average $V$
light curve (the symbols 1 and 2 are for curves $V_1$ and $V_2$).
The marked part of the light curves is also shown on a large scale
for better visibility of the error bars.} \label{fig:2}
\end{figure*}

The new folded light curves (Figs. \ref{fig:1},\ref{fig:2}) were built using the ephemeris \citepalias{dim12}
\begin{equation}\label{eq:1}
 HJD(\rm{Min})=2454417.382244(25)+0\fd1935980(11) \times E.
\end{equation}

The qualitative analysis of all photometric data led to several conclusions.

\begin{description}
\item[(i)] The out-of-eclipse light level changes from one orbital
cycle to another. Figure \ref{fig:1} illustrates this with the
data in filter $I$ obtained in two consecutive nights in January
2011. A similar behavior is visible in fig. 2 of
\citetalias{dim12} for the $V$ data.

\item[(ii)] The brightness of J0107 changes on time scales of
months. We found that the light levels during December 2011 both
into and out of the eclipse are lower by around $\Delta V \sim
0\fm10-0\fm15$ than those obtained in January 2011, while the
variability amplitudes are the same (Fig. \ref{fig:2}). The
brightness of both curves, $V_1$ from January and $V_2$ from
December 2011, is the same only at phases $\varphi \sim
0.10-0.15$.

\item[(iii)] The pre-eclipse hump is no permanent
feature of the light curve of J0107: it is clearly visible in
light curve $V_1$ but is lacking in curve $V_2$. As a result, the
light level after the eclipse is higher than before it in the
curve $V_2$ (Fig. \ref{fig:2}).
\end{description}

The measured depths (in mag) and widths (as a relative part of the
cycle) of the eclipse minimum in different filters (Table
\ref{tab:2}) imply a higher light contribution of the accretion
disk of J0107 at shorter wavelengths as well as a bigger size of
the secondary at longer wavelengths.

\begin{table}
\caption{Depths and widths of the eclipse}
\label{tab:2}
\centering
\begin{tabular}{c c c}
\hline\hline
filter& depth & width \\
\hline
$B$ & 3.0 & 0.060  \\
$V$ & 2.9 & 0.062 \\
$R$ & 2.4 & 0.063 \\
$I$ & 1.7 & 0.078 \\
\hline
\end{tabular}
\end{table}

The most distinguishable feature of J0107 is its deep eclipse (2.9
mag in $V$). The eclipse depths of the majority of the eclipsing
nova-like variables are $\leq$1 mag. The deepest eclipses belong
to the SW Sex stars. Only nine SW Sex stars in the list of
\citet{rodr07} have eclipse depths above 2.0 mag, which are however,
lower than that of J0107. Two SW Sex stars, DW UMa \citep{stan04}
and V1315 Aql \citep{pap09}, have shown deeper eclipses of 3.2 --
3.4 mag, but only once and only within  2 -- 3 days, whereas
normally their eclipse depths are below 2.0 mag. Therefore,
2MASS~J01074282+4845188 has the deepest permanent eclipse (at
least during 11 months) among the known eclipsing nova-like
variables.

\subsection{Spectral observations}

The spectra of the target (Table \ref{tab:1}) were obtained with
the 2-m RCC telescope that is equipped with a VersArray 512 CCD
camera, a focal reducer FoReRo-2, and a grism with 300 lines/mm.
The resolution of the spectra is 5.2 $\AA$/pixel and they cover
the range 5000 -- 7000 $\AA$.

\begin{figure}
\resizebox{0.8\hsize}{!}{\includegraphics{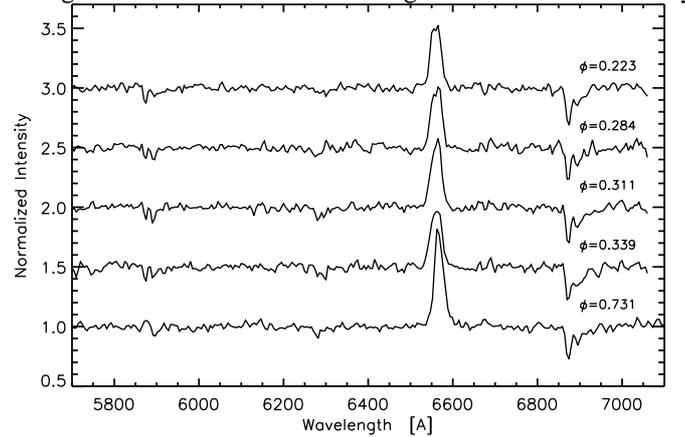}}
\caption{New low-resolution spectra of J0107.} \label{fig:3}
\end{figure}

The variability of the H$\alpha$ emission of J0107 is apparent in
the published four spectra of J0107 (fig. 3 of \citetalias{dim12})
as well as in the six new spectra (Fig. \ref{fig:3} and Table
\ref{tab:3}).

\begin{table*}
\caption{Parameters of the H$\alpha$ line of J0107}
\label{tab:3}
\centering
\begin{tabular}{l c c c c c c}
\hline\hline
Date          & phase &$\lambda_{c}$& FWHM & FWZI & EW  & Intensity \\
              &       &  [$\AA$]    &[$\AA$]&[$\AA$]&[$\AA$]& \\
\hline
2011 Feb 7    & 0.759 & 6566.2      & 24.9 & 68.0 & 13.8 & 1.55 \\
2011 Feb 7    & 0.783 & 6566.3      & 23.7 & 57.5 & 11.8 & 1.47 \\
2011 Feb 7    & 0.842 & 6565.4      & 26.8 & 73.2 & 11.4 & 1.43 \\
2011 Feb 7    & 0.860 & 6567.8      & 33.5 & 83.7 & 13.8 & 1.43 \\
2011 March 12 & 0.223 & 6559.6      & 28.3 & 72.7 & 16.7 & 1.51 \\
2011 March 12 & 0.284 & 6561.6      & 26.6 & 74.1 & 15.3 & 1.50 \\
2011 March 12 & 0.311 & 6561.0      & 28.2 & 79.4 & 16.9 & 1.55 \\
2011 March 12 & 0.339 & 6560.5      & 30.7 & 84.6 & 15.8 & 1.47 \\
2011 Nov 4    & 0.731 & 6567.1      & 24.7 & 76.8 & 20.2 & 1.74 \\
2011 Nov 4    & 0.749 & 6566.3      & 23.6 & 79.4 & 23.4 & 1.86 \\
\hline
\end{tabular}
\end{table*}

The qualitative analysis of the spectra led to several conclusions:

\begin{description}
\item[(a)] The intensity of the H$\alpha$ emission increases from
the beginning to the end of 2011 (Table \ref{tab:3}).

\item[(b)] There is a trend of the H$\alpha$ emission to be
considerably higher around phase 0.75.

\item[(c)] The H$\alpha$ emission line has predominantly
one-peaked profile.

\item[(d)] Even in the insufficient phase coverage of our spectral
observations, the H$\alpha$ line reveals an S-wave with an
amplitude of at least 270 km~s$^{-1}$ (Table \ref{tab:3}). This is
one of the most important criteria for an SW~Sex membership.

\item[(e)] The high values of the full width at zero intensity
(FWZI) of the H$\alpha$ line (Table \ref{tab:3}) are typical of
SW~Sex stars \citep{thor91}.
\end{description}

\section{Light curve solution of the Rozhen photometry of J0107}

\subsection{Model for the light curve solution}

To determine the parameters of the newly discovered CV J0107 we used the model of \citet[][hereafter
\textit{Model TK}]{khruz11} which is based on several assumptions.

\begin{enumerate}
\item It takes into account the effects of reflection, gravity
darkening, and linear limb darkening of the two stellar
components: the spherical white dwarf and the late secondary
filling its Roche-lobe, whose separation $a_{0}$ is
assumed to be a distance unit ($a_{0} = 1.0$).

\item The accretion disk around the white dwarf is slightly
elliptical ($e \leq 0.1$) and one of its foci coincides
with the white dwarf center (Fig. \ref{fig:4}). The equatorial
plane of the disk lies in the orbital plane of the system
\citep[see detailed description in][]{khruz00}.

\begin{figure}
\centering \resizebox{0.7\hsize}{!}{\includegraphics{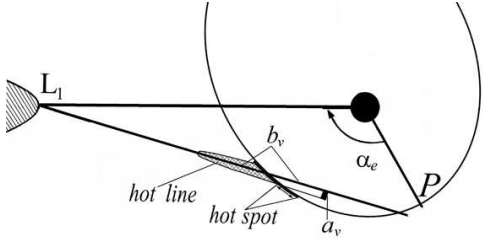}}
\caption{Configuration of a nova-like variable: the orbital plane
with the disk periastron $P$ and the optically opaque part of the
flow from the secondary (shaded area).} \label{fig:4}
\end{figure}

The disk is heated mainly by the gravitational energy released by
the accretion of matter on the white dwarf surface. As a result,
the temperature distribution through the disk is
\begin{equation}\label{eq:2}
T(r)=T_{\rm{in}}\Biggl(\frac{R_{\rm{in}}}{r}\Biggr)^{\alpha_{\rm{g}}}
,
\end{equation}

where $R_{\rm{in}}$ is the radius of the boundary layer of the
disk (usually it is assumed $R_{\rm{in}} \sim R_{\rm{wd}}$) while
$r$ is the distance from the center of the current differential
disk area to the center of the white dwarf. The range of the
parameter $\alpha_{\rm{g}}$ is $[0.10, 0.75]$: for low values of
$\alpha_{\rm{g}}$ the disk is bright, while for high values of
$\alpha_{\rm{g}}$ the disk emission is faint. The equilibrium
state of the disk corresponds to the highest value
$\alpha_{\rm{g}} = $0.75 \citep{shak73}.

\item The hot spot is heated by the shock wave arising from the
collision of the gas stream with the rotating disk
\citep{bis97,bis98,bis03,bis05}. The geometric thickness of the
disk is higher there (Fig. \ref{fig:5}). The shape of the hot spot
is described by a half ellipse on the lateral disk surface.

\begin{figure*}
\sidecaption
\includegraphics[width=0.65\hsize]{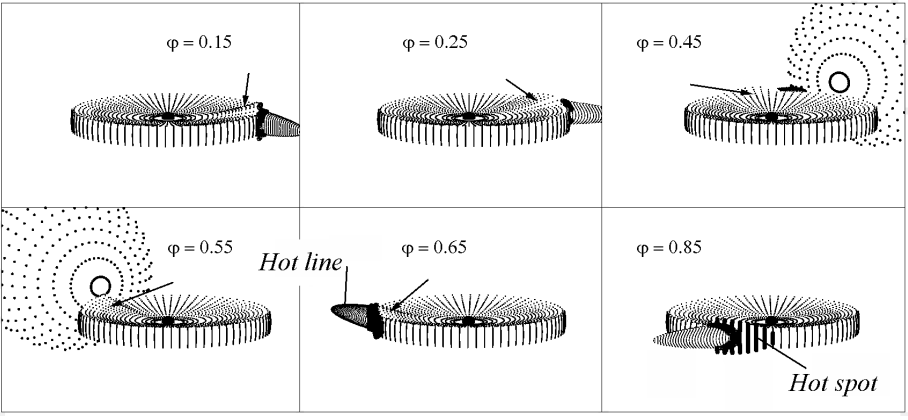}
\caption{3D configurations of a nova-like variable at different
orbital phases $\varphi$. The arrows show the disturbed area of
the inner disk surface while the thick dots indicate the heated
part of the hot line near its base.} \label{fig:5}
\end{figure*}

\item The opaque portion of the gas stream is reproduced by the
hot line. This feature is also heated by the shock wave and its
shape is assumed to be an elongated ellipsoid (Figs.
\ref{fig:4},\ref{fig:5}).

The emission from the hot spot and the hot line can explain the
pre-eclipse hump in the light curves of the nova-like variables.
Moreover, the hot line can cause a hump at the end of the eclipse
for certain orientations of its \textbf{biggest axis}.

\item To calculate the emission from the secondary non-spherical
star we took into account three effects.

The temperature distribution due to the gravity darkening is
\begin{equation}\label{eq:3}
 T(r_{\rm{i}})=T_2\Biggl[\frac{g(r_{\rm{i}})}{<g>}\Biggr]^\beta  ,
\end{equation}

where $g(r_{\rm{i}})$ is the gravitational acceleration at a
current point at distance $r_{\rm{i}}$ from the mass center of the
secondary, and $T_2$ is its temperature without heating.

The heating of the secondary from the high-temperature sources of
J0107 was calculated with the formula describing the heating of a
star from nearby hot (X-ray) component \citep{balog81}. For
CVs this formula is reduced to

\begin{equation}\label{eq:3a}
T^4_{\rm {h}}(r_{\rm {i}})=T^4(r_{\rm {i}}) + L_*\frac{\varepsilon\cos\alpha}{4\pi\sigma\rho^2}, \hspace{5mm} L_*=4\pi\sigma T^4_*R_{\rm {wd}}^2,
\end{equation}

where $\varepsilon$ is the \textbf{coefficient of processing} of
the high-temperature emission \citep[$\varepsilon \sim 0.5$ for
hot stars and $\varepsilon \sim 1.0$ for cool stars,][]{basko73};
$\alpha$ is the angle between the normal to the elementary heated
area $dS$ and the direction to the high-temperature source; $\rho$
is the distance between the high-temperature source and the heated
area $dS$. The \textit{Model TK} checks the visibility of the
white dwarf from each elementary area $dS$ of the secondary star
and then calculates its temperature according to eq.
(\ref{eq:3a}). For cataclysmic stars $T_*$ corresponds to
the temperature of the boundary layer of the accretion
disk because usually it is higher than $T_{\rm {wd}}$ (this is
amply fulfilled for our target, see below).

The limb-darkening is taken into account to calculate the emission
fluxes of the secondary

\begin{equation}\label{eq:3b}
F(\varphi)=\sum_{cos\gamma >0}{B_{\rm \lambda}}(T_{\rm {h}}(r_{\rm
{i}}))[1-u(\lambda ,T)+u(\lambda ,T) \cos \gamma] \cos \gamma dS ,
\end{equation}

where $u(\lambda,T)$ is the limb-darkening coefficient and $B_{\rm \lambda}(T_{\rm {h}}(r_{\rm {i}}))$
is the Planck function corresponding to the mean wavelength of the filter.

\end{enumerate}

The fitted parameters of \textit{Model TK} are

\begin{description}
\item[(a)] the mass ratio $q = M_{\rm{wd}}/M_{2}$ and orbital
inclination $i$;

\item[(b)] the secondary temperature $T_{2}$;

\item[(c)] the radius $R_{\rm{wd}}$ and temperature $T_{\rm{wd}}$
of the white dwarf;

\item[(d)] the parameters of the accretion disk: eccentricity $e$,
radius at the apoastron $R_{\rm{d}}$, azimuth of the disk
periastron $\alpha_{\rm{e}}$, temperature of the boundary layer
$T_{\rm{in}}$, temperature of the disk edge $T_{\rm{out}}$,
thickness of the disk outer edge (the height $z/2$ of the disk
edge above the orbital plane in terms of $a_{0}$ or the full
angular thickness $\beta_{\rm{d}}$ of the disk edge in degrees),
parameter $A$ of the paraboloid describing the internal disk
surface, parameter $a$ of the ellipsoid describing the external
disk surface, parameter $\alpha_{\rm{g}}$ determining the
temperature distribution across the disk;

\item[(e)] the parameters of the hot line: axis sizes
$a_{\rm{v}}$, $b_{\rm{v}}$, and $c_{\rm{v}}$, temperature
parameters $T_{\rm{w}}$ (windward temperature) and $T_{\rm{l}}$
(leeward temperature);

\item[(f)] the radius of the hot spot $R_{\rm{sp}}$ in units of
$a_{0}$ (the code calculates also the angle $\varphi_{\rm{sp}}$
(in degrees) between the axis of the hot line and the line
connecting the stellar components which depends on the parameters
of the disk and the hot line).
\end{description}

\subsection{Preliminary considerations and procedure of the light curve solution}

We present several considerations that allow us to narrow some
ranges of the unknown model parameters.

\begin{description}
\item[(a)] To decrease the range of the mass ratio we assumed that
J0107 belongs to the subtype SW~Sex of the nova-like variables
\citepalias{dim12}. According to the catalogs of \citet{ritt03}
and \citet{cher96}, the mass ratios of the SW~Sex systems are in
the range $q \sim 0.8-7.9$ with a peak near $q \sim 1-3$ (Table
\ref{tab:4}).

We were unable to use the eclipse width as an estimate of the mass
ratio $q = M_{\rm{wd}}/M_{2}$ and orbital inclination $i$ of the
system \citep{litt08} because the time resolution of the Rozhen
photometric observations of J0107 is insufficient.

\item[(b)] For the standard configuration of a nova-like variable
the disk radius is about half of the Roche lobe and its eclipse
(at least partial) is possible for orbital inclinations $i >
60\degr$. Despite the deep eclipse of J0107 we assumed the
initial condition $i > 40\degr$.

\item[(c)] The spectral types of the secondary components of the
SW~Sex stars are G -- M (Table \ref{tab:4}). Therefore we varied
the secondary temperature in the wide range $T_{2}\sim 3000-6000$
K.

Our spectra of J0107 do not allow us to constraint the secondary
temperature (it is too weak to show some features in our
low-resolution spectra). It is not reasonable either to use the
empirical relationship of \citet{warn95} between the orbital
periods of CVs and the absolute visual magnitudes of their
secondaries because it is based on only five nova-like stars.
\end{description}

\begin{table*}
\caption{Parameters of variable stars of SW~Sex subtype}
\label{tab:4}
\centering
\begin{tabular}{l c c c c c c}
\hline\hline
Name &$P_{\rm{orb}}$~[days]& Sp type & $q = M_{\rm{wd}}/M_2$ & $i$~[$\degr$] &$M_{\rm{wd}}/M_{\sun}$&$M_2/M_{\sun}$\\
\hline
SW Sex  & 0.134938& & 1.8& 79& 0.58& 0.33\\
BT Mon  & 0.333814 & G8V& 1.19& 82& 1.04& 0.87\\
V363 Aur& 0.321242 & G5--9V& 0.85& 60& 0.9& 1.06\\
& & & 1.12& 70& 0.86& 0.77\\
AC Cnc  & 0.300478 & K1--3V& 0.98& 75& 0.76& 0.77\\
V347 Pup& 0.231936 & M0.5V& 1.2--2.2& 84--87& 0.63--1.2& 0.52--0.55\\
RW Tri  & 0.231883 & M0V& 1.3& 70& 0.55& 0.35\\
& & & 0.76& 79& 0.44& 0.58\\
V1776 Cyg& 0.164739& & 1.6& 75& 0.6& 0.37\\
UU Aq   & 0.163805 & K7--M0V& 2.4--3.3& 71--83& 0.67--0.9& 0.2--0.4\\
LX Ser  & 0.158432 & M2V& 1.14--5& 75--90& 0.41& 0.36\\
V380 Oph& 0.154107 & & 1.6& 42& 0.58& 0.36\\
VZ Scl  & 0.144622 & & 0.7& 90& 1& 1.4\\
V1315 Aql& 0.139690& & 2.9& 82& 0.73& 0.30\\
DW UMa  & 0.136607 & M3--9& 2.6& 82& 0.77& 0.30\\
J0809+3814& 0.134038& & 3.3& 65& 1.0& 0.3\\
V348 Pup& 0.101839 & & 3.2& 81& 0.65& 0.20\\
EX Hya  & 0.068234 & M5--6V& 4.4--7.9& 78& 0.7--1.5& 0.1--0.2\\
\hline
\end{tabular}
\end{table*}

For the remaining fitted parameters we used the standard ranges:
$R_{\rm{d}}/ \xi \sim 0.3 - 0.9$ ($\xi$ is the distance between
the mass center of the white dwarf and $L_{1}$ whose value depends
on $q$ and is calculated with the code); $A \sim 3-7$;
$\alpha_{\rm{e}} \sim 0-90\degr$; $T_{\rm{in}} = \gamma
T_{\rm{wd}}$ where $\gamma \sim 1-7$ (the temperature of the
boundary layer $T_{\rm{in}}$ determines the temperature
distribution along the disk radius, the heating of the secondary
component and the heating of the inner disk regions near the outer
edge); $\alpha_{\rm{g}} \sim 0.1-0.75$; $R_{\rm{wd}}/ \xi \sim
0.001-0.05$ or $R_{\rm{wd}} \sim (0.001-0.03)a_{0}$; $T_{\rm{wd}}
\sim 10\,000-70\,000$ K; $T_{\rm{l}}, T_{\rm{w}} \sim 200-90\,000$
K; $R_{\rm{sp}}/R_{\rm{d}} \sim 0-0.25$; $a_{\rm{v}} \sim
(0.1-0.4)R_{\rm{d}}$, $b_{\rm{v}} \sim (0.2-0.9)R_{\rm{d}}$,
$c_{\rm{v}} \sim (1-1.8)z$.

To fit our data we used the code based on the \textit{Model TK}.
It calculates the light fluxes using the Planck energy
distribution for the corresponding central wavelength
$\lambda_{0}$ of the filters: 4400 $\AA$ for filter $B$, 5500
$\AA$ for filter $V$, 6960 $\AA$ for filter $R$, and 8800 $\AA$
for filter $I$ \citep{john65}. For the transfer from synthetic
light fluxes to magnitudes we used the value $F_{V}$=1.66 in
conventional units corresponding to $V = 15\fm09$. Using this
conventional unit is necessary because the Planck function defines
the energy flux through the unit area (1 cm$^{2}$) while the code
works with a distance unit $a_{0}$ whose value is unknown in
advance.

We used the tables of \citet{al78} and \citet{gry72} for the
values of the stellar limb-darkening coefficients $u(\lambda, T)$
corresponding to the wavelengths of the filters and stellar
temperatures and the value $u_{\rm{d}} = 2/3$ for the accretion
disk \citep{shak73}.

To search for the best fit the code uses the method of Nelder-Mead \citep{himm72}. An estimate of the fitting quality of the
theoretical and observed light curves is the sum of the residuals
\begin{equation}\label{eq:4}
 \chi^2=\sum^n_{j=1}\frac{(m_j^{\rm{theor}}-m_j^{\rm{obs}})^2}{\sigma^2_j},
\end{equation}
\noindent where $m_{j}^{\rm{theor}}$ and $m_{j}^{\rm{obs}}$ are the theoretical and observational magnitudes at the $j$-th orbital
phase, $\sigma_j$ is the dispersion of the $j$-th point and $n$ is the number of the points on the light curve.

It is known that an inverse problem solution is a
combination of model parameters for which the sum of the residuals
$\chi^2$ given by eq. (\ref{eq:4}) is below the corresponding
critical level of $\chi^2_{\eta, n}$ (whose tabulated values
increase both with increasing $n$ and decreasing $\eta$). Usually,
the value of the probability $\eta$ to reject the correct solution
is chosen to be 0.001.

But there are cases for which the values of $\chi^2$ of all
solutions are above the critical level of $\chi^2_{\eta, n}$.
These situations may arise from the small errors $\sigma_{i}$ of
the normal points of the observed light curves (see eq.
\ref{eq:4}) and/or from some imperfections of the model
(neglecting of some effects). Then it is impossible to determine
the ``true'' uncertainties of the model parameters. In these cases
it is appropriate to estimate the stability of the solution, i.e.,
the impact of changing a given parameter on the solution quality
\citep{cher93, abu08}. For this aim it is necessary to introduce
the ``conditional'' sum of the residuals $\chi^2 (+N \%)$. Its
value is higher by $N\%$ than the sum of the residuals of the
considered solution (usually $N\%=10\%$).

In order to find the global minimum of the residuals in
the parameter space we carried out the calculations in two stages.

Firstly, we searched for solutions for each filter separately and
for the whole considered ranges of the model parameters. Since the
dependence of the light curve shape on the mass ratio is weak,
this parameter was varied \textbf{by step of} $\Delta q = 0.1$. As
a result we determined the intersections of the ranges of the main
system parameters obtained for the different filters:
$q=[2.33-2.40]$; $i=[81-83]\degr$; $T_{2} = [4700-5200$]~K;
$R_{\rm{wd}}/ \xi = [0.01130-0.01145]$; $T_{\rm{wd}} =
[38\,000-41\,000]$~K; $A = [4.59-4.61]$. By subsequently varying
the main model parameters in these narrower ranges we obtained the
value: $q = 2.332$; $i = 81.4\degr$; $R_{\rm{wd}} = 0.0114 \xi$;
$T_{\rm{wd}} = 40\,000$~K; $T_{2} = 5090$~K.

At the second stage we fixed the values of the main model
parameters and varied the remaining model parameters that describe
the disk, the hot line, and the spot.

\begin{table*}
\caption{Parameters of the light curve solution of J0107}
\label{tab:5}
\centering
\begin{tabular}{l c c c c c c } \hline \hline
Parameters&Units&  $B$& $V_1$& $V_2$ & $R$ & $I$ \\
\hline{\it n} && 114 & 43 & 75 & 71 & 35 \\
\hline
$q = \frac{M_{\rm{wd}}}{M_2}$ &&\multicolumn{5}{c}{2.332$\pm$0.023} \\
$i$&\degr &\multicolumn{5}{c}{81.4$\pm$0.1} \\
$T_2$&K &\multicolumn{5}{c}{5090$\pm$20} \\
$<R_2>$& $a_0$ &\multicolumn{5}{c}{0.318$\pm$0.001} \\
$\xi$& $a_0$&\multicolumn{5}{c}{0.586$\pm$0.001} \\
\hline
&& \multicolumn{5}{c}{White dwarf} \\
\hline
$R_{\rm{wd}}$& $\xi$&\multicolumn{5}{c}{0.0113$\pm$0.0011} \\
$R_{\rm{wd}}$& $a_0$&\multicolumn{5}{c}{0.0066$\pm$0.0006} \\
$T_{\rm{wd}}$& K&\multicolumn{5}{c}{40000$\pm$600} \\
\hline
&&\multicolumn{5}{c}{Accretion disk} \\
\hline
$R_{\rm{d}}$& $\xi$& 0.362$\pm$0.009& 0.46$\pm$0.03& 0.381$\pm$0.005& 0.462$\pm$0.003& 0.50$\pm$0.04\\
$a$& $a_0$& 0.211$\pm$0.005& 0.27$\pm$0.02& 0.223$\pm$0.003& 0.270$\pm$0.001& 0.29$\pm$0.02\\
$A$&&  4.6$\pm$0.3& 4.6$\pm$0.9& 4.5$\pm$0.1&  4.60$\pm$0.03&  4.6$\pm$0.8\\
$z/2$& $a_0$& 0.0100$\pm$0.0002& 0.013$\pm$0.006& 0.010$\pm$0.001& 0.013$\pm$0.001& 0.014$\pm$0.001\\
$\beta$& \degr& 5.38$\pm$0.02& 5.4$\pm$0.9& 5.64$\pm$0.02& 5.43$\pm$0.02& 5.42$\pm$0.02 \\
$T_{\rm{in}}$& K& 44110$\pm$820 & 52470$\pm$7200 & 52350$\pm$800 & 58780$\pm$13100 & 58900$\pm$9550 \\
$T_{\rm{out}}$& K& 17300$\pm$110 & 23800$\pm$300 & 28600$\pm$110 & 31350$\pm$75 & 32150$\pm$140 \\
$\alpha_{\rm{g}}$&& 0.27$\pm$0.04& 0.22$\pm$0.04& 0.173$\pm$0.005& 0.17$\pm$0.01& 0.16$\pm$0.04 \\
\hline
&& \multicolumn{5}{c}{Hot line and hot spot} \\
\hline
$a_v$& $a_0$& 0.02$\pm$0.01& 0.09$\pm$0.04& 0.08$\pm$0.01&0.13$\pm$0.01 & 0.04$\pm$0.02 \\
$b_v$& $a_0$& 0.213$\pm$0.008& 0.30$\pm$0.03& 0.26$\pm$0.01& 0.36$\pm$0.01& 0.24$\pm$0.02 \\
$c_v$& $a_0$& 0.010$\pm$0.001& 0.02$\pm$0.01& 0.019$\pm$0.002& 0.023$\pm$0.001& 0.017$\pm$0.002 \\
$T_w$& K& 50000$\pm$2060& 77500$\pm$3770& 85120$\pm$3700& 67640$\pm$1700& 57110$\pm$2540 \\
$T_l$& K& 22750$\pm$3900& 1710$\pm$240& 1850$\pm$350& 2510$\pm$1600& 27180$\pm$2330 \\
$R_{\rm{sp}}$& $a_0$& 0.10$\pm$0.02& 0.2$\pm$0.1& 0.13$\pm$0.07& 0.24$\pm$0.02& 0.13$\pm$0.04 \\
$\varphi_{\rm{sp}}$& \degr& 8.5$\pm$0.5& 9$\pm$2& 6$\pm$2& 6.3$\pm$0.1&  14$\pm$4\\
\hline
 $\chi^2$&&  9177& 3419 &4607 & 3148 &1774  \\
 $\chi^2_{0.001,n}$&&  168.5& 78.2 &120.1 & 115 &66.9  \\
\hline
\end{tabular}
\end{table*}

\begin{figure*}
\resizebox{0.80\hsize}{!}{\includegraphics{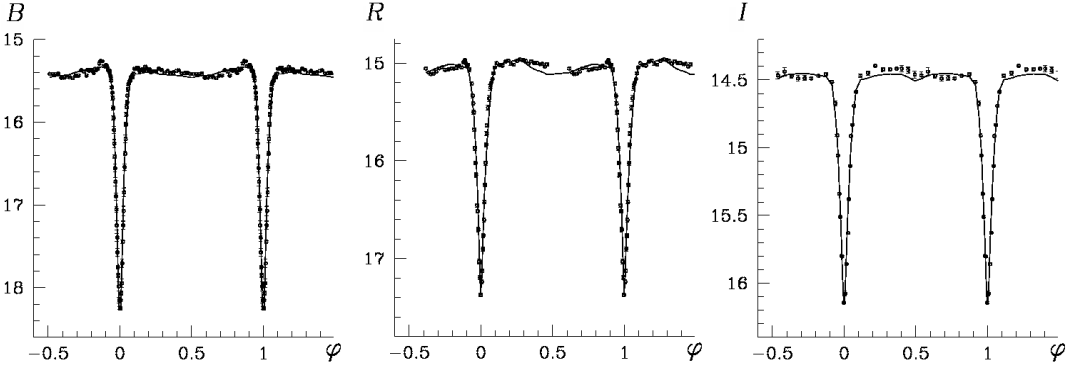}}
\caption{Synthetic light curves (continuous lines) and
observational data (points) in filters $BRI$.} \label{fig:6}
\end{figure*}

\begin{figure}
\resizebox{0.95\hsize}{!}{\includegraphics{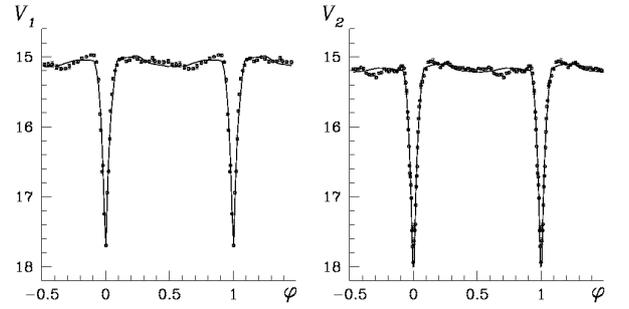}}
\caption{Synthetic light curves (continuous lines) in filter $V$
and observational data (points) for the two observational seasons
(January 2011 and December 2011).} \label{fig:7}
\end{figure}

Table \ref{tab:5} presents the results of the final light curve solution. 
The average synthetic light curves corresponding to the parameters
from Table \ref{tab:5} are shown in Figs. \ref{fig:6}, and \ref{fig:7}.

The uncertainties of the main parameters of the system $q$, $i$,
$T_2$, $T_{\rm{wd}}$, $R_{\rm{wd}}$ given in Table \ref{tab:5}
were determined as intersections of the errors of the
corresponding parameter for all five curves.

Owing to the considerably higher values of $\chi^2$  of our
solutions compared with the corresponding critical level of
$\chi^2_{0.001, n}$ (see the last rows of Table \ref{tab:5}), we
were unable to determine the ``true'' uncertainties of the fitted
parameters that describe the disk, the hot line, and the spot.
Instead we estimated the solution stability against the change of
these parameters around their derived values. Because there are
too many fitted parameters in our problem, we simply
varied each parameter $x_{k}$ in the range of its allowable
values, while the remaining parameters were fixed at the values
for which the lowest residual was obtained (corresponding
solution). The highest deviation $\Delta x_{k}$ from the solution
value $x_{k}$ for which the condition $\chi^2 < \chi^2(+10\%)$ was
fulfilled was assumed as the uncertainty of the fitted parameter
$x_{k}$ and is given in Table \ref{tab:5}.

\begin{figure}
\resizebox{0.95\hsize}{!}{\includegraphics{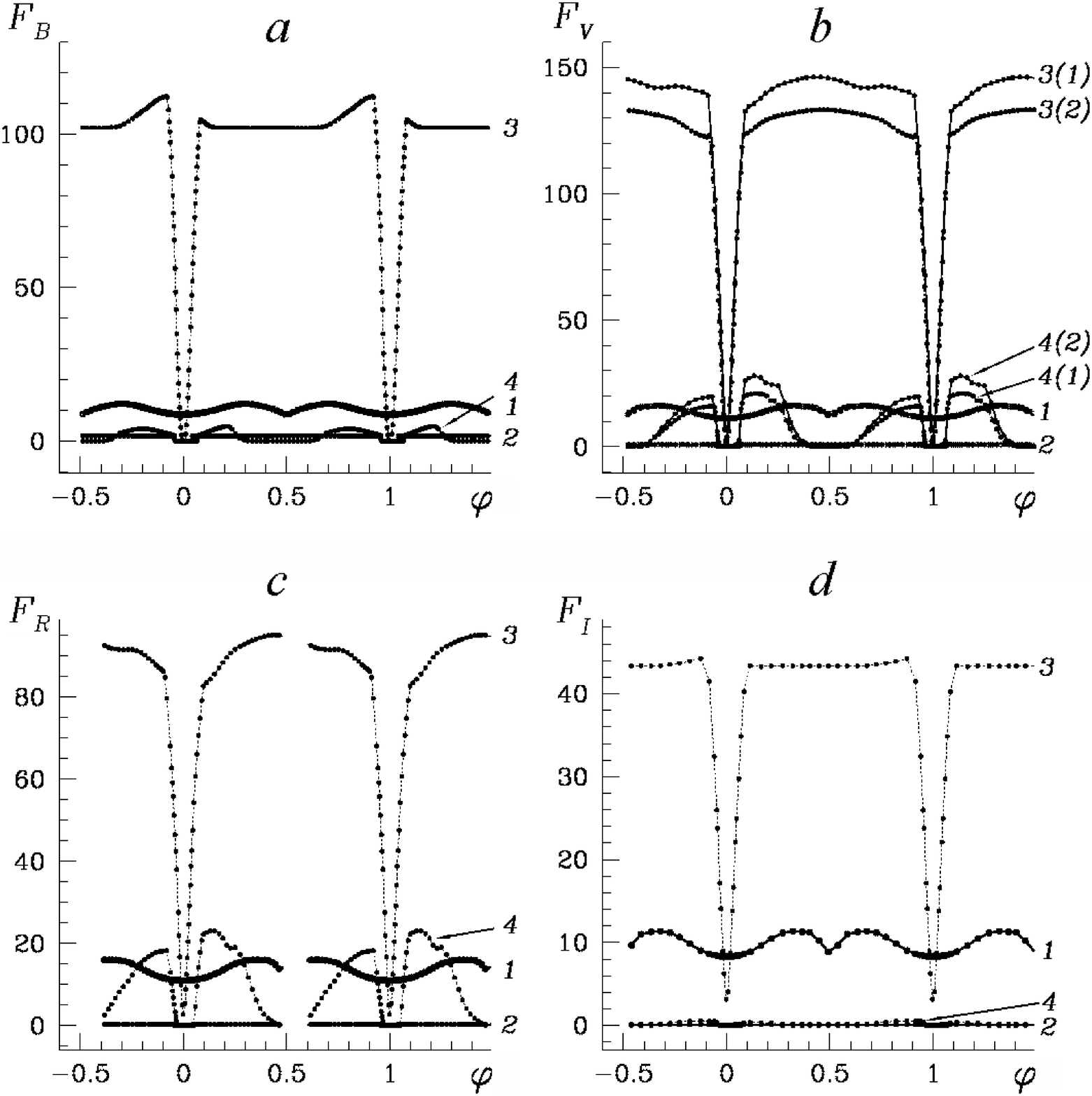}}
\caption{The relative light contributions of the secondary ({\it 1}), white dwarf ({\it 2}), accretion disk with the hot spot ({\it 3}) and hot-line ({\it 4}) in filters $B$
  ({\it a}), $V$ ({\it b}), $R$ ({\it c}), $I$ ({\it d}). These contributions are given separately for the two observational
seasons (January 2011 and December 2011) in filter $V$.}
\label{fig:8}
\end{figure}

Because the obtained ellipticity values $e$ of the accretion disk
were close to zero for all light curves ($ e \sim 0.00 - 0.02$),
the disk of J0107 could be assumed to be almost round. Then the
effect of parameter $\alpha_{e}$ on the solution is negligible.
For example, variations of $\alpha_{e}$ in the range from $0\degr$
to $180\degr$ for $e \sim 0.001$ lead to a change of $\chi^2$ up
to $0.5 - 1 \%$.

Our solution means that the eclipse of the accretion disk is total
only in filter $B$ (duration of the totality $\Delta \varphi_{B} =
0.0134$) and partial in the remaining bands (filters).

The disk luminosity is considerably higher (see Fig. \ref{fig:8})
than that of the remaining emission components (secondary star,
white dwarf, and hot line). Accordingly, the eclipse depth and the
light curve shape of J0107 mainly depend on the parameters of the
accretion disk (Figs. \ref{fig:6} -- \ref{fig:8}). 
The white dwarf contributes very little to the total luminosity
despite the high temperature because of its small radius. The hot
line emission causes some features on the out-of-eclipse part of
the light curves, which are more apparent in filters $V$ and $R$
(Fig. \ref{fig:8}).

\subsection{Analysis of the global parameters}

The analysis of the obtained system parameters of J0107 (Table \ref{tab:5}) led to several conclusions.

\begin{figure}
\resizebox{0.8\hsize}{!}{\includegraphics{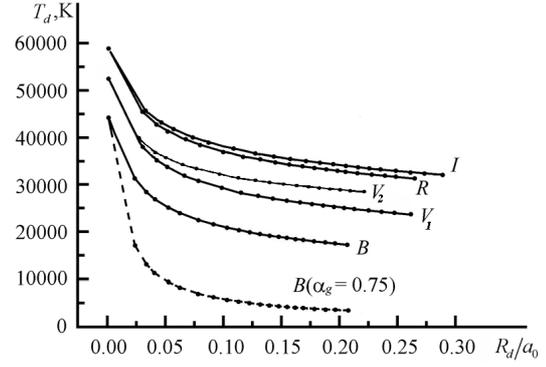}}
\caption{Temperature distributions along the disk radius of J0107 for the different spectral bands. The dashed line shows the distribution
for filter $B$ corresponding to the equilibrium value $\alpha_{\rm{g}} = 0.75$. The second point of each curve corresponds to the values
$R_{\rm{in}}$, $T_{\rm{in}}$.}
\label{fig:9}
\end{figure}

\begin{figure*}
\resizebox{0.99\hsize}{!}{\includegraphics{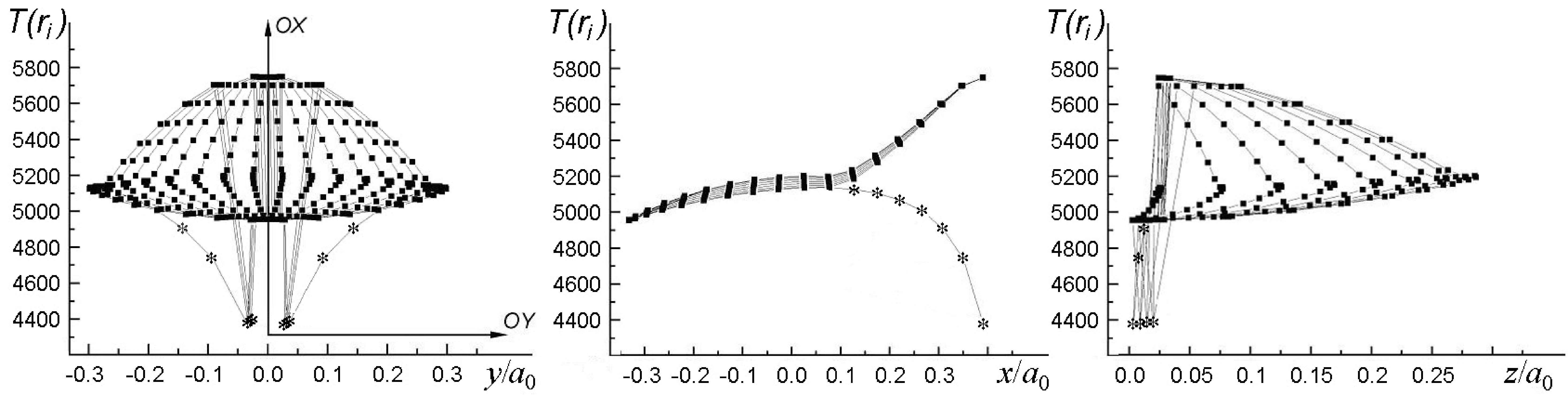}}
\caption{Temperature distribution of the secondary surface corresponding to the parameters of J0107 from Table \ref{tab:5}. The points near the
orbital plane which are shielded by the disk edge are marked by asterisks.}
\label{fig:10}
\end{figure*}

\begin{enumerate}
\item The obtained average radius of the secondary $\langle R_{2}
\rangle$=0.3181 (in units $a_{0}$) is close to the average radius
$R/a_{0} = 0.3085$ of the star filling its Roche lobe
obtained by the formulas of \citet{eggl83} for $q = 2.332$.

\item The radius of the accretion disk is relatively small,
comparable with that of the secondary component ($R_{\rm{d}} \sim
0.4 \xi$). The disk size is smallest in filter $B$, which
is the band of the deepest eclipse. The obtained value
$\beta_{\rm{d}} \sim 5\degr$ means that the disk of J0107 has a
relatively thick outer edge. The huge luminosity and relatively
small radius of the accretion disk mean relatively high
temperature.

\item We were able to reproduce the photometric data of J0107 well
with the values $q = M_{\rm{wd}}/M_{2} = 2.332$ and $i =
81.4\degr$. However, the main contributor to the light curve of
the system J0107 is the very bright accretion disk around the
white dwarf, not the stellar components that define the value of
$q$. Hence, radial velocity measurements of the target are
necessary to obtain a more reliable value of the mass ratio $q$.

\item The obtained value $T_{\rm{wd}}=40\,000$ K of J0107 is quite
high. The comparison with the reliable measurements of
$T_{\rm{wd}}$ in 43 CVs \citep{town09} reveals
that only two of them, DW UMa and MV Lyr, have hotter white dwarfs
(WDs) than J0107 and the $T_{\rm{wd}}$ of TT Ari is similar to
that of J0107. These three cases (DW UMa, MV Lyr and TT
Ari) are the only CVs of nova-like type from the whole sample.

\item The temperature $T_{\rm{in}}$ of the boundary layer of the
disk exceeds $T_{\rm{wd}}$ by factors of 1.1, 1.31, 1.46, and 1.47
in filters $B$, $V$, $R$, and $I$. The $T_{\rm{in}}$ values in all
filters are well above the standard temperatures for cataclysmic
variables at quiescence but are similar to those during their
outbursts. The determined outer edge temperatures $T_{\rm{out}}$
(Table \ref{tab:5}) of J0107 are also considerably higher (by
factors of 5--10) than the typical values $T_{\rm{out}} \sim
3000-5000$ K for the cataclysmic variables.

\item The obtained low values 0.16 -- 0.27 of $\alpha_{\rm{g}}$
(Table \ref{tab:5}) determine relatively flat temperature
distributions along the disk radius (Fig. \ref{fig:9}). These
values are also typical for CVs during outbursts
(when $\alpha_{\rm{g}}$ decreases up to $\sim 0.1$).

The obtained temperatures of the white dwarf and accretion disk
revealed that the high-temperature emission that heats the
secondary is generated mainly by the boundary layer of the disk.

\item Special attention was paid to the obtained high temperature
of 5090 K of the secondary of J0107. We attempted to obtain a good
light curve solution for temperatures appropriate for a red dwarf
secondary but they were far from the photometric data of J0107,
especially in the $B$ band. The solutions for $T_2$ values from
the range [3000, 4700] K corresponded to values of $\chi^2$ higher
by 30 -- 40 $\%$ whereas the values of the main parameters $q$,
$i$, $R_{\rm{wd}}$, and $T_{\rm{wd}}$ remained almost the same
(within 1 $\%$).

A review of Table \ref{tab:4} reveals that the high temperature of
the secondary of J0107 is more similar to long-period SW Sex star.
It was reasonable to assume that the high value of $T_2$ is caused
by the heating through the emission of both the high-temperature
white dwarf and the high-temperature accretion disk of J0107.

High temperatures of the secondaries of CVs have been established
during their outbursts. For instance, \citet{hess84} showed that
during the eruption of the dwarf-nova SS Cyg the temperature of
its K5 star increased to 16000 K on the side facing the white
dwarf, while \citet{rob86} found that this heating causes even an
underestimation of the measured radial velocities. Obviously, the
heating effect is essential for close binaries whose components
have quite different temperatures, as our target.

Our detailed calculations of the reflection effect of J0107
revealed that it decreases the contribution of the ellipsoidal effect
around phase $\varphi \sim 0.5$.
Otherwise, the synthetic light curve of the secondary would have a
deeper minimum at phase 0.5 than at phase 0.0, and
correspondingly, the total light curve of J0107 would have a dip
around phase 0.5. The very strong heating of the secondary in our
case is the reason that the light flux at phase $\varphi\sim 0.5$
is comparable with the fluxes from the secondary at the
quadratures. Indeed, the minimum of the secondary' light fluxes in
the phase range $\varphi\sim 0.4-0.6$ (see Fig. \ref{fig:8})
results from the covering (eclipse) of almost a quarter of the
hottest surface of the secondary star by the accretion disk.

Our tests with temperatures $T_{\rm{in}}$ lower than the value
corresponding to our $V$ light curve solution (Table \ref{tab:5})
led to a considerable worsening of the fit. For instance,
$T_{\rm{in}}=40000$ K would cause a decrease of the secondary' $V$
flux of up to 12 $\%$, while $T_{\rm{in}}=20000$ K would lead to a
decrease of up to 18 $\%$. Then the secondary' light curve would
become the same as the curve without any heating. Hence, each
underestimation of the secondary' heating causes undesired
deviations of the synthetic light curves from the observational
data.

Figure \ref{fig:10} shows the temperature distribution caused by
the heating effect on the secondary' surface.

One could expect that the prolonged heating of the secondary
surface in the part facing the white dwarf of J0107 would cause
stationary streams in the photosphere that strive to equalize the
temperatures of the neighboring volumes. Moreover, the turbulent
processes (vortical and acoustic) in the photosphere of the late
star (more intensive by two orders than in the hot star) also
favor the establishment of thermal equilibrium. Although the
energy transfer into the convective stellar layers is poorly
studied, some rough estimation of its time scale can be obtained
assuming that the thermalization of the absorbed heat occurs up to
a depth of several photospheric thicknesses. This estimate shows
that increasing the secondary surface temperature by 2000 K
through the turbulent transfer with acoustic speed would take
several days.

Consequently, the prolonged intensive heating from the hot white
dwarf and from the hot accretion disk as well as the turbulent
mixing of the photospheric gas of the secondary are probably the
main reasons for the higher temperature $T_2$ of J0107 compared to
those of the other nova-like variables with cooler white dwarfs.

\item The comparison of the light curve solutions of the $V$ light
curves from December 2011 and January 2011 revealed a decrease of
the disk radius, a decreasing thickness of the disk outer edge, an
increasing temperature at the disk outer edge, and a flatter
temperature distribution along the disk radius in December 2011
than that in January 2011 (Table \ref{tab:5}).
\end{enumerate}

\subsection{The non-steady emission of the accretion disk}

It is known that during the outbursts the emission of the CVs
increases considerably (by factor of hundreds) due to the
increased accretion. Moreover, the temperature of the CV disks
increases and the temperature distribution along the disk radius
becomes flatter \citep{djur96}, i.e., the emission of the CVs at
outbursts becomes non-steady.

\begin{figure}
\resizebox{0.8\hsize}{!}{\includegraphics{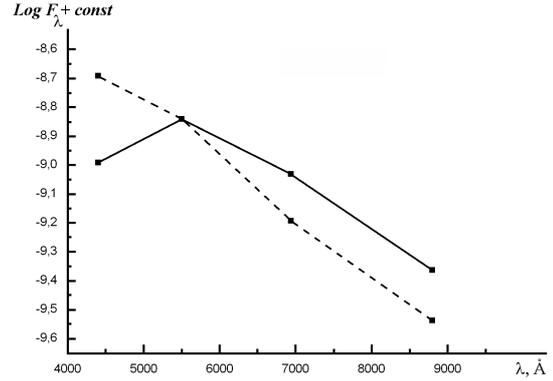}}
 \caption[]{Empirical dependence of the accretion disk emission on the wavelength (solid line) and the synthetic dependence of a disk
with black-body emission whose parameters are equal to the
solution for filter $V$ (dashed line). The points refer to the
average fluxes at each filter for the phase range 0.2 -- 0.7
(without effects of disk eclipse and hot spot).} \label{fig:11}
\end{figure}

Our light curve solution revealed that the parameter
$\alpha_{\rm{g}}$ of J0107 was $\sim 0.2$ for all filters, i.e.,
far from the steady state value 0.75. Moreover, the empirical
dependence of the emission of J0107 on the wavelength differs from
the black-body type (Fig. \ref{fig:11}), which is the second
appearance of the non-steady emission of its accretion
disk.

The theoretical fluxes in the red part of the spectrum ($R$ and
$I$ bands), which correspond to the light curve solution in filter
$V$ ($T_{\rm{in}} = 52\,468$ K and $\alpha_{\rm{g}} = 0.2156$),
are half as strong as the observed fluxes (Fig. \ref{fig:11}).
Higher temperatures of the boundary layer $T_{\rm{in}} \sim
58\,000-59\,000$ K and a flatter temperature distribution along
the disk ($\alpha_{\rm{g}} \sim 0.16-0.17$) than those in the $V$
filter were necessary to provide the observed disk emission flux
and the relative contribution of the disk to the total system
luminosity in the red part of the spectrum. A larger disk radius
would not be able to give the necessary effect because it would
lead to an eclipse shape different from the observed one. The
situation for filter $B$ is the reverse: the observed disk
flux is about two times smaller than the theoretical flux for the
values of $T_{\rm{in}}$ and $\alpha_{\rm{g}}$ corresponding to the
light curve solution in filter $V$ (Fig. \ref{fig:11}).

With obtained value $T_{\rm{wd}}=40\,000$ K and assuming an
average mass 0.7 M$_{\sun}$ of the white dwarf (see Table
\ref{tab:4}), we obtained an approximate estimation of the mass
accretion rate of J0107 $\dot{M}=8\times 10^{-9}$
M$_{\sun}$~yr$^{-1}$ \citep[according to formula (2)
of][]{town09}. This high value can explain the high luminosity of
its disk.

The derived exceptionally high disk temperatures, the quite low
values of the parameter $\alpha_{\rm{g}}$, and the high mass
accretion rate of J0107 are typical for CVs at
eruption. But they are permanent for J0107 which means that its
emission is permanently non-steady. We consider the
similarities of the nova-like variable J0107 to CVs
at eruption as additional arguments to the statement that the
nova-like stars are permanently in an active state.

\subsection{The newly discovered nova-like variable J0107 versus the prototype UX~UMa}

The empirical data and their analysis led us to the conclusion
that the nova-like J0107 revealed many peculiarities. To exhibit
them we compared the global characteristics of the newly
discovered nova-like star J0107 with those of the well-known
nova-like eclipsing binary UX~UMa whose photometry was fitted with
the same model \citep{kjurk06,khruz07}. We found the following
significant differences.

\begin{description}
\item[(a)] The temperature of 40\,000 K of the white dwarf of
J0107 is considerably higher than the corresponding value 27\,000
K of UX~UMa. \item[(b)] The temperature $T_{\rm{in}}$ of the
boundary layer of the accretion disk of J0107 is higher than
50\,000 K, while that of UX~UMa is below 30\,000 K. \item[(c)] The
temperature of 5090 K of the secondary of J0107 is considerably
higher than the corresponding value 3400 K of UX~UMa. \item[(d)]
The characteristic temperatures of the hot line of J0107 are
considerably higher (by a factor of 4 -- 6) than those of the hot
line of UX~UMa.
\end{description}

We assume that these (mainly) temperature differences cause the
different emission characteristics of the compared targets, which
are:

\begin{description}
\item[(i)] The empirical dependence of the emission on the
wavelength of J0107 differs considerably from the black-body type,
while that of UX~UMa is almost of black-body type \citep[see fig.
16 in][]{kjurk06}; \item[(ii)] The $\alpha_{\rm{g}}$ value of
UX~UMa is about 0.6, i.e., near to the steady-state value of 0.75,
while the $\alpha_{\rm{g}}$ value of J0107 is about 0.2, which is
far from the equilibrium value.
 \end{description}

Hence, there is considerable deviation of the emission of the
bright accretion disk of J0107 from the steady type, while the
emission of UX~UMa is almost steady. We suggest that the
non-steady emission of the accretion disk of J0107 may be
attributed to the low viscosity of the disk matter caused by its
unusual high temperature. Analyses of the emission of more targets
of this type are necessary to check this assumption.

An additional difference between J0107 and UX~UMa is that the
H$\alpha$ emission line of UX~UMa has a two-peaked profile at most
phases, while that of J0107 is rather single-peaked.

\subsection{SW~Sex classification of J0107}

The inherent features of SW~Sex stars continuously evolve with
increasing statistics but the same is true for the other subtypes
of CVs. Moreover, some characteristics of the different subtypes
overlap. This means that the rich world of CVs does not permit
firm boundaries between the various types.

In order to check if the new photometric and spectral data and
their modeling support or disclaim the initial supposition about
the belonging of J0107 to the SW~Sex class we used the
characteristics of this subtype as defined by \citet{thor91},
\citet{rodr07}, and \citet{schm09}.

\begin{description}
\item[(a)] J0107 shows the typical V-shaped eclipse of SW~Sex
stars, maybe one of the first main signs of this subtype.
Recently, an eclipse is not considered a mandatory criterion of
the SW~Sex stars because it is biased toward high-inclination
systems. Today, at least 13 out of a total of 35 confirmed SW~Sex
stars do not display eclipses \citep{rodr07}. \item[(b)] Our
target poses high-velocity emission line wings, typical of SW~Sex
stars (Table \ref{tab:3}). \item[(c)] The one-peaked profile of
the H$\alpha$ emission line is typical of SW~Sex stars
\citep{thor91}. \item[(d)] J0107 shows a very weak \ion{He}{I}
6678 emission line, which is typical of SW~Sex stars. \item[(e)]
Our low-resolution spectra revealed an S-wave for the H$\alpha$
line with an amplitude of about 270 km~s$^{-1}$ (Table
\ref{tab:3}). \item[(f)] J0107 shows high luminosity and a hot
white dwarf, typical of SW~Sex subtype. \item[(g)] The obtained
high value $8\times 10^{-9} M_{\sun}$~yr$^{-1}$ of the mass
accretion rate of J0107 is another confirmation that it belongs to
the SW~Sex stars.
\end{description}

Owing to the low-resolution of our spectra, their insufficient
phase coverage, and insufficient $\lambda$ coverage, we are unable
to confirm (or disprove) the additional spectral signs of SW~Sex
stars \citep{rodr07}: narrow absorption features in the Balmer and
\ion{He}{I} lines near the inferior conjunction of the white
dwarf; orbital phase offsets of the radial velocity curves with
respect to the photometric ephemeris; \ion{He}{II} 4686 emission;
line flaring (change of the emission line flux and velocity with
periods of about 10 -- 20 min) similar to that in the intermediate
polar CVs.

The value $P_{\rm{orb}}=4.65$ h of J0107 is slightly higher than
the period range 3.0 -- 4.5 h of the most of the known SW~Sex
stars \citep{rodr07}. But there are members with considerably
longer periods \citep[see our Table \ref{tab:3} and table 6
of][]{rodr07}.

It was established recently that half of the nova-likes known to
undergo VY~Scl faint states (system brightness drops by up to 4 --
5 mag for months) are SW~Sex stars \citep{rodr07}. Until today,
J0107 has not shown VY~Scl-type light decreasing.

According to \citet{schm09} an S-wave and central absorption are
the main characteristics of the SW~Sex stars, and correspondingly,
they should be searched for with spectral observations. Reviewing
the available information for the SW~Sex variables, we established
that especially important characteristics of these stars are the
high temperatures of their white dwarfs and accretion disks. To
exhibit these, photometric data are necessary (also in UV and
X-ray regions).

\subsection{Evolution state}

After establishing that the most of the nova-like stars in the 3
-- 4 h period range (just above the period gap) are of SW~Sex
subtype \citep{rodr07,schm09}, it was reasonable to assume that
because of the angular momentum loss each CV with a longer orbital
period will fall into the 3 -- 4 h period range and will turn into
a SW~Sex star, i.e., the SW~Sex phenomenon can be
considered as an evolutionary stage of the CV population.

The important evolution role of the SW~Sex stars requires future
precise observations and modeling to study and to explain their
peculiarities.

The unusual high luminosities of the SW~Sex stars are not properly
explained. High mass-transfer \citep{liv94} or an additional light
source \citep[such as nuclear burning,][]{hon01} were proposed as
plausible reasons.

Recently, it was assumed that the accretion at a very high rate
$\dot{M}$ is the most likely explanation for the high luminosity
of the SW~Sex stars. It may be caused by enhanced mass transfer
from the donor star due to heating of its inner face by a very hot
white dwarf. Indeed, just the nova-like CVs have the hottest white
dwarfs (with effective temperatures of up to 50\,000 K) found in
any CVs \citep{town09}. The results of our light curve solution of
J0107 completely confirm all previous theoretical suggestions.

The extremely high accretion rates of the SW~Sex stars
\citep[around $5\times 10^{-9} M_{\sun}$~yr$^{-1}$,][]{town03}
explain their permanent high state but why SW~Sex stars have the
highest mass-transfer rates it has yet to be satisfactorily
explained. This is a cornerstone because it is obvious that the
mass-transfer rates govern the evolution of CVs.

The high values of $T_{\rm{wd}}$ and $\dot{M}$ are assumed to be
common characteristics of VY~Scl and SW~Sex stars. It was
suggested that these two subtypes represent a phase in the CV
evolution at which the binaries evolve into semi-detached
configurations with a short peak of the mass transfer
\citep{rodr07}. Some SW~Sex stars do undergo VY~Scl-type light
decreasing \citep[for instance DW~UMa, classified even as VY~Scl
subtype by][]{town09}, but there are dwarf novae \citep[like
HT~Cas and RX~And,][]{sch02} that also show VY~Scl-type light
decreasing. New data and analyses are necessary to derive
additional criteria for the sub-classification of the CVs and to arrange them chronologically in the evolution
scheme.

\section{Conclusions}

The light curve solution of the nova-like variable 2MASS J01074282+4845188 led to the following conclusions.

\begin{enumerate}
\item The observed excessively deep light minimum is reproduced by
the eclipse of the very bright accretion disk whose contribution
is much higher than those of the other light sources.

\item The obtained unusual high temperatures of the disk of J0107
and the low values of the parameter $\alpha_{\rm{g}}$ are typical
of Cvs during outburst. These similarities of
our new nova-like variable and the CVs at eruption
might be additional arguments for the statement that the nova-like
stars are permanently in an active state.

\item The flat temperature distribution along the disk radius and
the deviation of the energy distribution from the black-body law
mean that the disk emission of J0107 is non-steady. This
result can be attributed to the low viscosity of the disk matter
due to its unusual high temperature.

\item The primary of J0107 is one of the hottest white dwarfs in
CVs.

\item The temperature 5090 K of the secondary of J0107 is quite
high and more appropriate to a long-period SW~Sex star. Our
detailed analysis revealed that it could be mainly attributed to
the intense heating from the hot white dwarf and the hot
accretion disk.

\item The high mass accretion rate $\dot{M} = 8\times 10^{-9}
M_{\sun}$~yr$^{-1}$ of J0107 implies that it is in a very
short evolutionary stage.

\item The broad and single-peaked H$\alpha$ profile of J0107 and
the S-wave surely confirm its classification as an SW~Sex star.

\item The main differences between the global parameters 
of J0107 and those of the well-known nova-like eclipsing binary
UX~UMa are the considerably higher temperatures of all components
of the J0107 configuration (white dwarf, accretion disk,
secondary, and hot line) compared to those of UX~UMa.
\end{enumerate}

Initially, the newly discovered nova-like eclipsing star J0107
attracted our attention because of its excessively deep eclipse.
Furthermore, it turned out that this is the deepest permanent
eclipse (at least during 11 months) among the known nova-like
stars. Finally, the light curve solution showed that some global
parameters of J0107 have quite extreme values (especially the
temperatures). We conclude that the established peculiarities of
2MASS~J01074282+4845188 make it an interesting target for
follow-up spectral observations and observations in UV and X-ray
bands.

\begin{acknowledgements}
This study is supported by funds of the projects DO 02-362, DO
02-85 and DDVU 02/40-2010 of the Bulgarian National Science Fund,
as well as by the Russian Foundation for Basic Research (project
nos. 09-02-00225, 11-02-00258) and the Program of State Support to
Leading Scientific Schools of the Russian Federation
(NSh-2374.2012.2). DD and DK gratefully acknowledge observing
grant support from the Institute of Astronomy and Rozhen National
Astronomical Observatory, Bulgarian Academy of Sciences. The
authors are very grateful to the anonymous referee for the useful
recommendations and to K.V. Bychkov from the Moscow University for
the discussion about the time-scale of heating of the secondary
star. This research makes use of the SIMBAD and Vizier data bases,
operated at CDS, Strasbourg, France, and NASA’s Astrophysics
Data System Abstract Service.
\end{acknowledgements}

\bibliographystyle{aa}
\bibliography{AA_2012_20385.bib}

\end{document}